\documentclass[twocolumn,showpacs,preprintnumbers,amsmath,amssymb]{revtex4}
\usepackage{amssymb}
\usepackage{amsfonts}
\usepackage{mathrsfs}
\usepackage{amsmath}
\usepackage{graphicx}
\usepackage{dcolumn}
\usepackage{bm}
\usepackage{hyperref}

\begin{document}
\title{Practical decoy state measurement-device-independent quantum key distribution}
\author{Shi-Hai Sun$^1$\footnote{Email:shsun@nudt.edu.cn}, Ming Gao$^2$, Chun-Yan Li$^1$, and Lin-Mei Liang\footnote{Email:nmliang@nudt.edu.cn}$^{1,3}$}
\affiliation{$^1$Department of Physics, National University of Defense
Technology, Changsha 410073, P.R.China\\
$^2$State Key Laboratory of Mathematical Engineering and Advanced Computing, Zhengzhou 450002, P.R.China\\
$^3$State Key Laboratory of High Performance Computing, National University of Defense Technology, Changsha 410073, P.R.China}
\begin{abstract}
Measurement-device-independent quantum key distribution (MDI-QKD) is immune to all the detection attacks, thus when it is combined with the decoy state method, the final key is unconditional secure, even if practical weak coherent source are used by Alice and Bob. However, until now, the analysis of decoy state MDI-QKD with weak coherent source is incomplete. In this paper, we derive, with only vacuum+weak decoy state, some tight formulas to estimate the lower bound of yield and the upper bound of error rate for the fraction of signals in which both Alice and Bob send single photon pulse to the untrusted third party Charlie. The numerical simulations show that our method with only vacuum+weak decoy state can asymptotically approach to the theoretical limit of the infinite number of decoy states. Furthermore, the statistical fluctuation due to the finite length of date is also considered based on the standard statistical analysis.
\end{abstract}


\maketitle
\emph{Introduction-} Quantum key distribution (QKD), such as the BB84 protocol \cite{BB84}, admits two remote parties, known as Alice and Bob, to share unconditional security key, which is guaranteed by the quantum mechanics and has been proved in theory \cite{Lo99,Shor00,GLLP04}. However, the setups used in the practical system are imperfect, which will leave some loopholes for Eve to spy the secret key. In fact, some potential quantum hacking strategies have been discovered by exploiting the imperfection of practical setups, such as the passive faraday mirror attack \cite{Sun11}, blinding attack \cite{Lydersen10}, time-shift attack \cite{Zhao08}, and so on \cite{Gisin06,Makarov06,Fung07}. Therefore, the legitimate parties must carefully reexamine their practical system to close all the loopholes, when they use this system in practical situations.

In order to close the gap between the theory and practice, some approaches have been proposed. The first one is trying to characterize the practical system fully and considered all the side-channel existed in the practical system. Although some potential loopholes have been discovered and then closed by using this approach, it can not find all the loopholes existed in the practical system, since, theoretically speaking, the number of loophole is infinite. The second approach is trying to establish full device-independent (DI-) QKD system \cite{Acin07,Pironio09}. The DI-QKD can guarantee the unconditional security of the practical system without knowing the detailed information of the practical setups of Alice and Bob. However, this approach is impractical within current technology, since it requires that the legitimate parties have single photon detectors with near unit detection efficiency.

In stead of full DI-QKD, recently, Lo \emph{et al.} proposed a novel scheme called measurement-device-independent (MDI-) QKD \cite{Lo12}, in which both Alice and Bob send pulse to an untrusted third party, called Charlie. Charlie performs the Bell state measurement (BSM) and tells her results to Alice and Bob, then Alice and Bob can use this information to distill secret key. Since the detection party can be fully controlled by the eavesdropper (Eve), this scheme is immune to all the detector attacks. Thus the legitimate parties just need to ensure that the source is secret, then the total QKD system is secret. In fact, this condition can be satisfied in practical situations, since the source is relatively simple and can be fully characterized.

Although the MDI-QKD has been demonstrated in experiments \cite{Rubenok12,Liu12}, and some modified schemes for fiber-based system have bee proposed \cite{Ma12,Tamaki12}, it is not completely device-independent. It requires the source of Alice and Bob is perfect, for example the pulse sent by Alice and Bob should be single photon state. However, within current technology, the weak coherent state is often used due to the lack of a feasible single photon source, which will send multi-photon pulses with nonzero probability and suffer from the photon-number-splitting (PNS) attack \cite{Huttner95,Brassard00}. Luckily, the same problem is also faced for the regular BB84 protocol with the weak coherent state, and the decoy state method \cite{Hwang03,Lo05,Wang05,Ma05} has been proposed to efficiently estimate the contribution of single photon pulse. Thus the decoy state method can also be introduced to the MDI-QKD to close the loophole of the multi-photon pulses.

However, the analysis for the decoy state MDI-QKD is different from the regular decoy state QKD for the regular BB84 protocol \cite{Hwang03,Lo05,Wang05,Ma05}. Recently, the security of the decoy state MDI-QKD has been considered by many researchers \cite{Lo12,Ma12,Ma-Fung12,Song12,Wang12}. However, there still exists some disadvantages for theirs results. In Ref.\cite{Lo12}, Lo \emph{et al.} analyze the security of decoy state MDI-QKD assuming infinitely long data and infinitely many decoy states, which is impractical due to the limited resource in practical situations. In Ref.\cite{Ma12,Ma-Fung12,Song12}, the authors considered the effect of the finite-size data and a finite number of decoy states, but their analysis has two disadvantages: first, the authors estimate the contribution of single-photon pulses by solving the nonlinear minimization problem, but not giving general formulas liking the regular decoy state QKD; second, four states (vacuum+two-weak decoy state) are needed to close to the asymptotic limit of infinitely decoy states. Therefore, a more stringent security bound and the general theory of decoy state MDI-QKD is imperative.

In this paper, we discuss the decoy state MDI-QKD with vacuum+weak decoy state, in which both Alice and Bob use three kinds of state with different intensity (one signal state, one decoy state and one vacuum state). Then we derive general formulas to estimate the yield $Y_{11}$ and error rate $e_{11}$ for the fraction of signals in which both Alice and Bob send single photon pulse to Charlie. The numerical simulations show that our formulas are very tight, and our vacuum+weak decoy state method asymptotically approaches to the theoretical limit of the infinite decoy state method.

\emph{Protocol-} In this paper, we consider the following decoy state MDI-QKD protocol \cite{Lo12,Ma12,Ma-Fung12}:

(1)Alice randomly generates three kind of pulses with different intensity, the signal state with a intensity $\mu_2$, the decoy state with a intensity $\mu_1$ and the vacuum state with a intensity $\mu_0\equiv 0$. Without loss of generality, we assume that $\mu_2>\mu_1>0$. For each pulse, Alice randomly chooses her basis from $\{x,z\}$ and bit from $\{0,1\}$. Then she modulates her information on each pulse and sends it to Charlie, which can be fully controlled by Eve. At the same time, Bob performs the same processing as Alice, and the intensities of Bob's pulse are noted as $\nu_2$, $\nu_1$ and $\nu_0\equiv0$ ($\nu_2>\nu_1>0$) for signal state, decoy state and vacuum state, respectively.

(2)Charlie performs BSM, and tells her measurement results to Alice and Bob through a public channel. Then Alice and Bob compare their basis for each pulse. If they use the same basis and Charlie has a successional BSM event, they keep this bit as raw key.

(3)For each case that Alice's intensity is $\mu_i$, Bob's intensity is $\nu_j$ and the basis is $\omega=x,z$,  Alice and Bob estimate the parameters of channel, including the total gain $Q_{\mu_i\nu_j}^\omega$, the total error rate $E_{\mu_i\nu_j}^\omega$, and the yield (error rate) of both Alice and Bob send single photon pulse, noted as $Y_{11}^\omega$ ($e_{11}^\omega$). With these parameters, Alice and Bob can estimate the final key rate, which is given by \cite{Lo12,Ma-Fung12}
\begin{equation}\label{key_rate}
R\geq \mu_2\nu_2 e^{-\mu_2-\nu_2}Y_{11}^z[1-H(e_{11}^x)]-Q_{\mu_2\nu_2}^z f H(E_{\mu_2\nu_2}^z),
\end{equation}
where $f$ is the error correction inefficiency, $H(x)=-x \log_2(x)-(1-x)\log_2(1-x)$ is the binary Shannon entropy function. Note that $Q_{\mu_2\nu_2}^z$ and $E_{\mu_2\nu_2}^z$ are directly measured in experiment, thus Alice and Bob need to estimate the lower bound of $Y_{11}^z$ and upper bound of $e_{11}^x$ to maximize her key rate. The main contribution of this paper is that we give two tight formulas to estimate $Y_{11}^z$ and $e_{11}^x$ with only vacuum+decoy state. Here we assume that only the signal states of Alice and Bob, $\mu_2$ and $\nu_2$, are used to distill the secret key. The decoy states are used to estimate the parameters of channel.

Note that, when the phase of pulse sent by Alice and Bob is totally randomized, the quantum channel can be considered as a photon-number channel model \cite{Lo05,Ma-Fung12}, and the state of Alice and Bob is
$\rho_\mu =\sum_{n=0}^\infty \frac{\mu^n}{n!}e^{-\mu}|n\rangle\langle n|$,
where $\mu=\{\mu_i,\nu_j|i,j=0,1,2\}$. Thus, the total gain and error rate of Alice's intensity $\mu_i$ and Bob's intensity $\nu_j$ can be written as \cite{Ma-Fung12}
\begin{equation}\label{gain_error}
\begin{split}
Q_{\mu_i\nu_j}^\omega&=\sum_{n,m=0}^\infty \frac{\mu_i^n\nu_j^m}{n!m!}e^{-\mu_i-\nu_j}Y_{nm}^\omega\\
E_{\mu_i\nu_j}^\omega Q_{\mu_i\nu_j}^\omega&=\sum_{n,m=0}^\infty \frac{\mu_i^n\nu_j^m}{n!m!}e^{-\mu_i-\nu_j}Y_{nm}^\omega e_{nm}^\omega,
\end{split}
\end{equation}
where $Y_{nm}^\omega$ ($e_{nm}^\omega$) is the yield (error rate) when Alice sends $n-$photon pulse, Bob sends $m-$photon pulse, and the basis $\omega$ is used by them. Obviously, according to Eq.\ref{gain_error}, if infinite decoy states are used, Alice and Bob can exactly obtain $Y_{11}^z$ and $e_{11}^x$. However, the resource is finite in practical situations, thus only finite decoy state can be used by the legitimate parties. In the following, we give two tight formulas to bound these parameters, which are the main contributions of this paper. The numerical simulations show that our formulas with only vacuum+weak decoy state can asymptotically approach the theoretical limit of infinite decoy states.

\emph{The lower bound of $Y_{11}^\omega$-} Note that the expression of Eq.\ref{gain_error} is independent on $\omega$, thus when there is on ambiguity, we neglect the superscript $\omega$ in the following of this paper.
Then the total gain $Q_{\mu_i\nu_j}$ can be written as
\begin{equation}\label{gain}
\begin{split}
&e^{\mu_i+\nu_j}Q_{\mu_i\nu_j}=\sum_{n,m=0}^\infty \frac{\mu_i^n\nu_j^m}{n!m!}Y_{nm}\\
=&\sum_{m=0}^\infty \frac{\nu_j^m}{m!}Y_{0m}+\mu_i(Y_{10}+\nu_j Y_{11}+\sum_{m=2}^\infty \frac{\nu_j^m}{m!}Y_{1m})\\
&+\sum_{n=2}^\infty \frac{\mu_i^n}{n!}(Y_{n0}+\nu_j Y_{n1}+\sum_{m=2}^\infty \frac{\nu_j^m}{m!}Y_{nm})\\
=&e^{\nu_j} Q_{0\nu_j}+e^{\mu_i} Q_{\mu_i0}-Q_{00}+\mu_i\nu_j Y_{11}+h(\mu_i,\nu_j),
\end{split}
\end{equation}
where
\begin{equation}
h(\mu_i,\nu_j)= \sum_{m=2}^\infty \frac{\mu_i\nu_j^m}{m!}Y_{1m}
+\sum_{n=2}^\infty\frac{\mu_i^n\nu_j}{n!} Y_{n1}
+\sum_{n,m=2}^\infty \frac{\mu_i^n\nu_j^m}{n!m!}Y_{nm}.
\end{equation}

Thus we will have
\begin{equation}
\begin{split}
&e^{\mu_2+\nu_2}Q_{\mu_2\nu_2}-e^{\mu_1+\nu_1}Q_{\mu_1\nu_1}\\
=&g_1 +(\mu_2\nu_2-\mu_1\nu_1)Y_{11} +\sum_{m=2}^\infty \frac{\mu_2 \nu_2^m-\mu_1\nu_1^m}{m!}Y_{1m}\\
&+\sum_{n=2}^\infty\frac{\mu_2^n\nu_2-\mu_1^n\nu_1}{n!} Y_{n1}  +\sum_{n,m=2}^\infty\frac{\mu_2^n\nu_2^m-\mu_1^n\nu_1^m}{n!m!}Y_{nm} \\
\geq& g_1 + (\mu_2\nu_2-\mu_1\nu_1)Y_{11} +a \sum_{m=2}^\infty \frac{\mu_2\nu_1^m+\mu_1 \nu_2^m}{m!}Y_{1m}\\
&+ b \sum_{n=2}^\infty\frac{\mu_2^n\nu_1+\mu_1^n\nu_2}{n!} Y_{n1} +c \sum_{n,m=2}^\infty \frac{\mu_2^n\nu_1^m+\mu_1^n\nu_2^m}{n!m!}Y_{nm} \\
\geq&g_1 + (\mu_2\nu_2-\mu_1\nu_1)Y_{11} +\alpha[h(\mu_2,\nu_1)+h(\mu_1,\nu_2)]\\
=&g_1 +g_2 +g_3 -(\mu_1\nu_1-\mu_2\nu_2+\alpha\mu_2\nu_1+\alpha\mu_1\nu_2)Y_{11},
\end{split}
\end{equation}
where we use the fact that for any $n,m\geq2$, the following inequalities always hold, which are given by
\begin{equation}
\begin{split}
\frac{\mu_2\nu_2^m-\mu_1\nu_1^m}{\mu_2\nu_1^m+\mu_1\nu_2^m}\geq \frac{\mu_2\nu_2^2-\mu_1\nu_1^2}{\mu_2\nu_1^2+\mu_1\nu_2^2}\equiv a \geq 0,\\
\frac{\mu_2^n\nu_2-\mu_1^n\nu_1}{\mu_2^n\nu_1+\mu_1^n\nu_2}\geq \frac{\mu_2^2\nu_2-\mu_1^2\nu_1}{\mu_2^2\nu_1+\mu_1^2\nu_2}\equiv b \geq 0,\\
\frac{\mu_2^n\nu_2^m-\mu_1^n\nu_1^m}{\mu_2^n\nu_1^m+\mu_1^n\nu_2^m}\geq \frac{\mu_2^2\nu_2^2-\mu_1^2\nu_1^2}{\mu_2^2\nu_1^2+\mu_1^2\nu_2^2}\equiv c\geq 0.
\end{split}
\end{equation}
And $\alpha=\min\{a,b,c\}$. Here $g_1$, $g_2$ and $g_3$ are defined as
\begin{equation}\label{3g}
\begin{split}
g_1&=e^{\nu_2}Q_{0\nu_2}+e^{\mu_2}Q_{\mu_20}-e^{\nu_1}Q_{0\nu_1}-e^{\mu_1}Q_{\mu_10},\\
g_2&=\alpha (e^{\mu_2+\nu_1}Q_{\mu_2\nu_1}-e^{\nu_1}Q_{0\nu_1}-e^{\mu_2}Q_{\mu_20}+Q_{00}),\\
g_3&=\alpha (e^{\mu_1+\nu_2}Q_{\mu_1\nu_2}-e^{\nu_2}Q_{0\nu_2}-e^{\mu_1}Q_{\mu_10}+Q_{00}).
\end{split}
\end{equation}
It is easy to check that for any $\alpha$, $\mu_1\nu_1-\mu_2\nu_2+\alpha\mu_2\nu_1+\alpha\mu_1\nu_2>0$ always holds. And note that the expression of equations from Eq.\ref{gain} to Eq.\ref{3g} are the same for both z-basis and x-basis. Thus the lower bound of $Y_{11}^\omega$
is given by
\begin{equation}\label{Y11}
Y_{11}^\omega\geq \underline{Y_{11}^\omega}\equiv\frac{g_1^\omega+g_2^\omega+g_3^\omega-
e^{\mu_2+\nu_2}Q_{\mu_2\nu_2}^\omega+e^{\mu_1+\nu_1}Q_{\mu_1\nu_1}^\omega}{\mu_1\nu_1-\mu_2\nu_2+\alpha\mu_2\nu_1+\alpha\mu_1\nu_2}.
\end{equation}
where $\omega=z,x$.

\emph{The upper bound of $e_{11}^\omega$-}
According to Eq.\ref{gain_error} and Eq.\ref{gain}, we have
\begin{equation}
\begin{split}
&e^{\mu_1+\nu_1}Q_{\mu_1\nu_1}E_{\mu_1\nu_1}
=g_4+\mu_1\nu_1 Y_{11}e_{11}+h'(\mu_1,\nu_1),
\end{split}
\end{equation}
where
\begin{equation}\label{g4}
\begin{split}
g_4=&e^{\nu_1} Q_{0\nu_1}E_{0\nu_1}+e^{\mu_1} Q_{\mu_10}E_{\mu_10}-Q_{00}E_{00},\\
h'(\mu_1,\nu_1)
&=\sum_{m=2}^\infty \frac{\mu_1\nu_1^m}{m!}Y_{1m}e_{1m}
+\sum_{n=2}^\infty\frac{\mu_1^n\nu_1}{n!} Y_{n1}e_{n1}\\
&+\sum_{n,m=2}^\infty \frac{\mu_1^n\nu_1^m}{n!m!}Y_{nm}e_{nm}.
\end{split}
\end{equation}
Obviously, $h'(\mu_1,\nu_1)\geq 0$, thus the upper bound of $e_{11}^\omega$ can be written as
\begin{equation}\label{e11}
e_{11}^\omega\leq \overline{e_{11}^\omega}
\equiv \frac{e^{\mu_1+\nu_1}Q_{\mu_1\nu_1}^\omega E_{\mu_1\nu_1}^\omega-
g_4^\omega}{\mu_1\nu_1 \underline{Y_{11}^\omega}},
\end{equation}
where $\omega=z,x$, and $\underline{Y_{11}^\omega}$ and $g_4$ are given by Eq.\ref{Y11} and Eq.\ref{g4} respectively.

\emph{Numerical Simulation-} Note that when Eve is absent, the total gains and error rates of Alice's intensity $\mu_i$ and Bob's intensity $\nu_j$ are given by \cite{Ma12,Ma-Fung12}
\begin{equation}\label{gain_error_simul}
\begin{split}
Q_{\mu_i\nu_j}^x&=2y^2[1+2y^2-4yI_0(s)+I_0(2s)],\\
Q_{\mu_i\nu_j}^x E_{\mu_i\nu_j}^x&=e_0Q_{\mu_i\nu_j}^x-2(e_0-e_d)y^2[I_0(2s)-1],\\
Q_{\mu_i\nu_j}^z&=Q_C+Q_E,\\
Q_{\mu_i\nu_j}^z E_{\mu_i\nu_j}^z&=e_dQ_C+(1-e_d)Q_E,
\end{split}
\end{equation}
where
\begin{equation}
\begin{split}
Q_C&=2(1-P_d)^2e^{-\mu'/2}[1-(1-P_d)e^{-\eta_a\mu_i/2}]\\
&\times[1-(1-P_d)e^{-\eta_b\nu_j/2}]\\
Q_E&=2P_d(1-P_d)^2e^{-\mu'/2}[I_0(2s)-(1-P_d)e^{-\mu'/2}].
\end{split}
\end{equation}
And $I_0(s)$ is the modified Bessel function of the first kind, $e_d$ is the misalignment-error probability, $e_0=1/2$ is the error rate of background, $P_d$ is the dark count of single photon detector, $\eta_a$ ($\eta_b$) is the transmission of Alice (Bob), and
$\mu'=\eta_a \mu_i+\eta_b\nu_j$, $s=\sqrt{\eta_a\mu_i\eta_b\nu_j}/2$, $y=(1-P_d)e^{\mu'/4}$.

\begin{figure}
\scalebox{1}{\includegraphics[width=8cm]{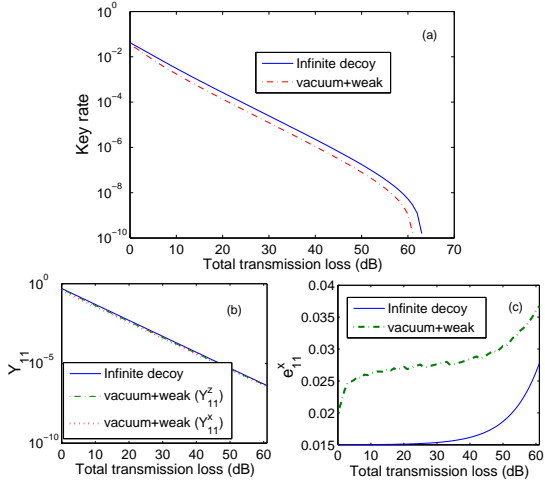}}
\caption{\label{fig:key_rate}(Color online) The key rate of decoy state MDI-QKD. The solid line is obtained for the infinite decoy state method, in which the exactly $Y_{11}^z$ and $e_{11}^x$ are known. The dot-dashed line is obtained for our vacuum+weak decoy state method, in which the lower bond of $Y_{11}^z$ and the upper bound of $e_{11}^x$ are given by Eq.\ref{Y11} and Eq.\ref{e11}, respectively. The key rate is maximized by optimizing the intensity of pulse, which is shown in Fig.\ref{fig:uv_opt}. The same parameters as Ref.\cite{Ma-Fung12} are used in our simulations, which are $e_d=1.5\%$, $P_d=3\times10^{-6}$, $f=1.16$.}
\end{figure}

\begin{figure}
\scalebox{1}{\includegraphics[width=8cm]{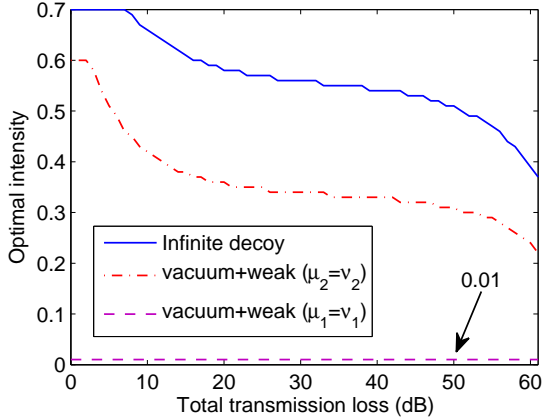}}
\caption{\label{fig:uv_opt}(Color online) The optimal intensity for signal state and decoy state to maximize the key rate. The optimal intensity is obtained by researching the intensity of signal state ($\mu_2$ and $\nu_2$)and decoy state ($\mu_1$ and $\nu_1$) from 0.01 to 0.6 with a step 0.01. In the simulations, we assume that $\eta_a=\eta_b$, $\mu_2=\nu_2$ and $\mu_1=\nu_1$. And other parameters are the same as Fig.\ref{fig:key_rate}.}
\end{figure}

Submitting Eq.\ref{gain_error_simul} into Eq.\ref{Y11} and Eq.\ref{e11}, we can estimate the lower bound of yield $Y_{11}^z$ and upper bound of error rate $e_{11}^x$ when both Alice and Bob send single photon state. The estimated parameters of $Y_{11}$ and $e_{11}$ are shown in Fig.\ref{fig:key_rate}(b) and (c) respectively, which clearly shows that our vacuum+weak decoy state method is very close to the asymptotic limit of the infinite decoy state method. Then, with these parameters, we can estimate the key rate, which is shown in Fig.\ref{fig:key_rate}(a). It clearly shows that the key rate with our method is also very close to the asymptotic limit of the infinite decoy state method. Note that the key rate is maximized by optimizing the intensity of signal state and decoy state. The optimal intensity for our method and infinite decoy state method are shown in Fig.\ref{fig:uv_opt}. It shows clearly that the optimal signal intensity is the order of $O(1)$, which is the same as the regular decoy state.

\begin{table}
\caption{\label{tab:Table 1} The comparing between our method and Ma's method. We assume that $\eta_a=\eta_b=0.1$. Here we directly take the results of Ma's method from Ref.\cite{Ma-Fung12}.}
\tabcolsep0.15in
\doublerulesep2pt
\begin{tabular}{cccc}
\hline\hline
Parameters &our method &Ma's method with\\
&& vacuum+weak \\
&($\mu_2=\nu_2=0.36)$&($\mu_2=\nu_2=0.5$)\\
\hline
$Y_{11}^z$ &$4.1967\times10^{-3}$ &$4.6043\times10^{-3}$\\
\hline
$e_{11}^x$ &2.7241\% &10.2126\%\\
\hline
R &$1.3548\times10^{-4}$ &$6.8877\times10^{-5}$\\
\hline\hline
\end{tabular}
\end{table}

Furthermore, our method can perform better than the method proposed by Ma \emph{et al.} \cite{Ma-Fung12}, which estimated the contribution of single photon state, $Y_{11}^z$ and $e_{11}^x$, by solving the nonlinear minimization problem. The results are listed in the Table \ref{tab:Table 1}. It clearly shows that the key rate estimated by our method is larger than that of Ma's method.

\emph{Statistical Fluctuation-} In practical situations, the length of raw key is also finite, which will induce statistical fluctuation for the parameters estimation. In this section, we considered the affect of finite length of raw key based on the standard statistical analysis \cite{Ma05,Ma-Fung12}, in which the lower bound and upper bound of experimental results, $Q_{\mu_i\nu_j}^\omega$ and $E_{\mu_i\nu_j}^\omega$, are given by
\begin{equation}
\begin{split}
\underline{Q_{\mu_i\nu_j}^\omega}&\leq Q_{\mu_i\nu_j}^\omega \leq \overline{Q_{\mu_i\nu_j}^\omega},\\
\underline{Q_{\mu_i\nu_j}^\omega E_{\mu_i\nu_j}^\omega}&\leq Q_{\mu_i\nu_j}^\omega E_{\mu_i\nu_j}^\omega\leq \overline{Q_{\mu_i\nu_j}^\omega E_{\mu_i\nu_j}^\omega},
\end{split}
\end{equation}
where $\underline{Q_{\mu_i\nu_j}^\omega}=Q_{\mu_i\nu_j}^\omega(1-\beta_q)$, $\overline{Q_{\mu_i\nu_j}^\omega}=Q_{\mu_i\nu_j}^\omega (1+\beta_q)$, $\underline{Q_{\mu_i\nu_j}^\omega E_{\mu_i\nu_j}^\omega}=Q_{\mu_i\nu_j}^\omega E_{\mu_i\nu_j}^\omega(1-\beta_{eq})$, $\overline{Q_{\mu_i\nu_j}^\omega E_{\mu_i\nu_j}^\omega}=Q_{\mu_i\nu_j}^\omega E_{\mu_i\nu_j}^\omega(1+\beta_{eq})$, and $\beta_q=n_\alpha/\sqrt{N_{\mu_i\nu_j}^\omega Q_{\mu_i\nu_j}^\omega}$, $\beta_{eq}=n_\alpha/\sqrt{N_{\mu_i\nu_j}^\omega Q_{\mu_i\nu_j}^\omega E_{\mu_i\nu_j}^\omega}$. Here $N_{\mu_i\nu_j}^\omega$ is the length of pulse of Alice's intensity $\mu_i$, Bob's intensity $\nu_j$ and $\omega$ basis. $n_\alpha$ is the standard deviations, which is related to the failure probability of the security analysis. For example, if $n_\alpha=5$, the failure probability is $5.73\times10^{-7}$ \cite{Ma-Fung12}. Thus the lower bound of $Y_{11}^\omega$ and upper bound of $e_{11}^\omega$, which are given by Eq.\ref{Y11} and Eq.\ref{e11}, should be rewritten as
\begin{equation}\label{Y11_sta}
\begin{split}
Y_{11}^\omega\geq& \underline{\underline{Y_{11}^\omega}}\equiv\frac{\underline{g_1^\omega}+\underline{g_2^\omega}+\underline{g_3^\omega}-
e^{\mu_2+\nu_2}\overline{Q_{\mu_2\nu_2}^\omega}+e^{\mu_1+\nu_1}\underline{Q_{\mu_1\nu_1}^\omega}}{\mu_1\nu_1-
\mu_2\nu_2+\alpha\mu_2\nu_1+\alpha\mu_1\nu_2},\\
e_{11}^\omega\leq& \overline{\overline{e_{11}^\omega}}\equiv\frac{e^{\mu_1+\nu_1}\overline{Q_{\mu_1\nu_1}^\omega E_{\mu_1\nu_1}^\omega}-\underline{g_4^\omega}}{\mu_1\nu_1 \underline{\underline{Y_{11}^\omega}}},
\end{split}
\end{equation}
where $\underline{g_k^\omega} (k=1,2,3)$ and $\underline{g_4^\omega}$ are given by Eq.\ref{3g} and Eq.\ref{g4}.

\begin{figure}
\scalebox{1}{\includegraphics[width=8cm]{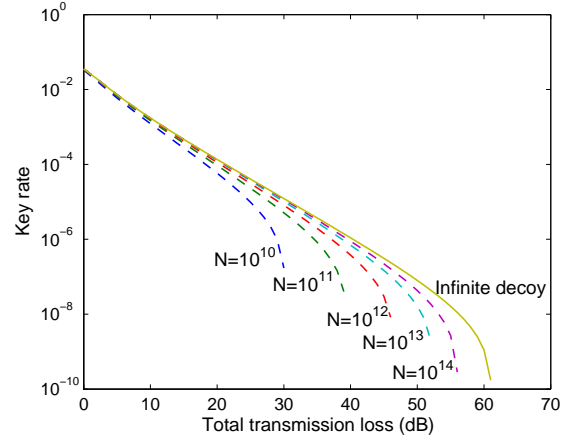}}
\caption{\label{fig:key_rate_fluc}(Color online) The key rate of decoy state MDI-QKD with statistical fluctuation. The solid line is obtained for the infinite decoy state method with infinite length of data. The dashed lines are obtained for our vacuum+weak decoy state with different length of data. In the simulations, we assume that five standard deviation ($n_\alpha=5$) is used. N is the length of data.}
\end{figure}

Submitting the equations above into Eq.\ref{key_rate}, we can estimate the secret key rate with finite length of data, which is shown in Fig.\ref{fig:key_rate_fluc}. It clearly shows that the finite length of raw key will obviously compromise the secret key rate. In the simulations, we assume the standard deviation is $n_\alpha=5$ and the length of data is same for each pair of intensities of Alice and Bob.

\begin{figure}
\scalebox{1}{\includegraphics[width=\columnwidth]{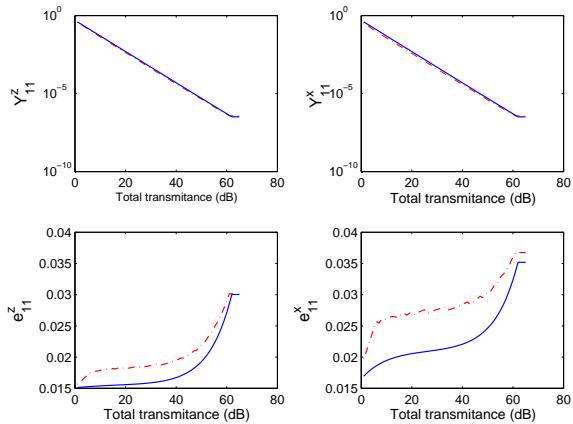}}
\caption{\label{fig:com_fig1}The estimated yield and error rate of single photon pulse for z-basis and x-basis. The solid lines are obtained by using Wang's Eq.(16) for yield and the corrected Eq.(17) for error rate \cite{Wang12}. The dashed lines are obtained by our method. The estimated values are optimized by searching the optimal intensity of Alice and Bob, and other parameters are same as this paper.}
\end{figure}

\begin{figure}
\scalebox{1}{\includegraphics[width=\columnwidth]{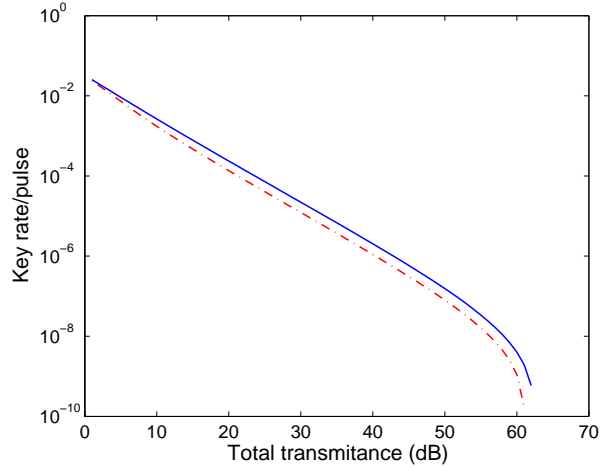}}
\caption{\label{fig:com_fig2} The key rate of our method and Wang's method. The solid lines are obtained by Wang's method with the corrected formula of Eq.(17). The dashed lines are obtained by our method.}
\end{figure}

\emph{Discussion-} In the first version of this paper, we claim that ``\emph{In Ref.\cite{Wang12}, Wang presents general formulas for the decoy state MDI-QKD with three intensity states (vacuum+weak decoy state), but their formulas are very relaxant, and no secret key can be generated when these foumulas are applied.}'' We acknowledge that this conclusion is obtained by directly applying Eq.(16) and Eq.(17) of Ref.\cite{Wang12} to estimate the yield and error rate of single photon pulse, and we only compare the final secret key rate in the simulations, but do not individually analyze and compare the estimated values of yield (Eq.(16)) and error rate (Eq.(17)). Recently, Wang updates a new version of his paper, in which a corrected formula is given to estimate the error rate of single photon pulse. Taking this correction, we recalculate the key rate of Wang's method, and find that the key rate obtained by his method is a bit higher than that of our method, thus we withdraw the claim and show these new simulation results in this revised version, which are given by Fig.\ref{fig:com_fig1} and Fig.\ref{fig:com_fig2}. The two figures clearly show that both Wang's method and our method can give a better estimation of yield and error rate of single photon pulse, and Wang's Eq.(16) \cite{Wang12}, the formula for the single-photon yield, is actually a bit better than that of our Eq.(\ref{Y11}).

\emph{Conclusions-} The MDI-QKD can exclude all the detection loopholes in practical situations, and when it is combined with the decoy state method, the final key generated by the MDI-QKD is unconditional security, even the practical weak coherent sources are used by Alice and Bob. However, the security of decoy state MDI-QKD is incomplete. In this paper, we discuss the decoy state MDI-QKD with vacuum+weak decoy state, in which both Alice and Bob use three kinds of state with different intensity (one signal state, one decoy state and one vacuum state). Then we derive general formulas to estimate the yield and error rate for the fraction of signals in which both Alice and Bob send single photon pulse to Charlie. The numerical simulations show that our formulas are very tight, and our method with vacuum+weak decoy state method asymptotically approaches to the theoretical limit of the general decoy state method (with an infinite number of decoy states).

\emph{Acknowledgement-} This work is supported by the National Natural Science Foundation of China, Grant No. 61072071, Grant No. 11204377, and Grant No. U1204602. L.M. Liang is supported by the Program for NCET. M. Gao is supported by National High-Tech Program of China, Grant No. 2011AA010803. S.H. Sun is supported by the Fund of Innovation, Graduate School of NUDT, Grant No. B100203.


\end{document}